\documentclass[aps,pre,twocolumn,groupedaddress,showpacs,floatfix]{revtex4}
\usepackage{amsmath}
\usepackage{amssymb}
\usepackage{graphicx}

\begin{document}

\title{Characterizing Polygonality in Biological Structures}

\author{Luciano da Fontoura Costa} 
\affiliation{Instituto de F\'{\i}sica de S\~ao Carlos. 
Universidade of S\~ ao Paulo, S\~{a}o Carlos,
SP, PO Box 369, 13560-970, 
phone +55 16 3373 9858,FAX +55 16 3371
3616, Brazil, luciano@if.sc.usp.br}

\author{Fernando Rocha and Silene Maria Ara\'ujo de Lima}
\affiliation{Centro de Ci\^encias Biol\'ogicas, \\
Departamento de Fisiologia, Universidade Federal do Par\'a, \\
Campus Universit\'ario do Guam\'a, Rua Augusto Corr\^ea 01, \\
CEP 66075-000, Bel\'em, Par\'a, \\
Brazil}

\date{9th Oct 2005}

\begin{abstract}   

Several systems involve spatial arrangements of elements such as
molecules or cells, the characterization of which bears important
implications to biological and physical investigations.  Traditional
approaches to quantify spatial order and regularity have relied on
nearest neighbor distances or the number of sides of cells. The
current work shows that superior features can be achieved by
considering angular regularity. Voronoi tessellations are obtained for
each basic element and the angular regularity is then estimated from
the differences between the angles defined by adjacent cells and a
reference angle. In case this angle is 60 degrees, the measurement
quantifies the hexagonality of the system.  Other reference angles can
be considered in order to quantify other types of spatial
symmetries. The performance of the angular regularity is compared with
other measurements including the conformity ratio (based on nearest
neighbor distances) and the number of sides of the cells, confirming
its improved sensitivity and discrimination power. The superior
performance of the haxagonality measurement is illustrated also with
respect to a real application concerning the characterization of
retinal mosaics.

\end{abstract}


\maketitle

\section{Introduction}

Several important properties of biological systems are directly
related and even determined by the spatial distribution of their
constituent elements.  For instance, the distribution of ganglion
cells through the mammals retina is known to accurately sample the
visual field with just the right amount of overlap.  Similarly, the
adjacency of cells is important for cell signaling, while the spatial
distribution of cells actively expressing some specific gene is
immediately related to tissue and organ formation.  Other examples
where the spatial distribution of the elements is critical involve
the spread of pathological agents such as viruses and bacteria as well
as the spatial arrangement of diseased cells and lesions.

Typically, spatial order in biological systems involves
structured/symmetric arrangements of points such as in hexagonal
systems.  These structures can be understood as regular lattices,
characterized by fixed spacing and angles between the constituent
points.  Even in the cases where the boundaries between the elements
are not available, they can be obtained from the Voronoi tessellation
considering the original points as seeds (e.g.~\cite{Okabe:2000}).
The ubiquity of polygonal organization in biological systems is
related to special packing and physical properties allowed by ordered
systems (e.g.~\cite{Ziman:1979}).  For instance, it is known that the
hexagon is the regular polygon with maximum number of sides which can
be used to tile the plane.  The immediate advantage of such a regular
tiling is the maximization of the number of neighbors of each element,
with immediate implications for cell signaling~\cite{Escriba:1995},
resolution and isotropy in sensory acquisition (e.g. photoreceptors,
retinal neurons and omatidea), structural properties (e.g. choral,
cell membrane, striate muscle fibers, and other structures), to name
but a few cases.  Interesting combinations of pentagons and hexagons
are also often found in three-dimensional biological structures such
as pollen, viruses and radiolaria structures.

Because spatial regularity plays such a key role in defining specific
structural and functional properties of biological systems at micro,
meso and macroscopic spatial scales, it becomes of paramount
importance to quantify in an objective and accurate way the degree of
spatial organization of such systems.  Among the previous works aimed
at automated quantification of spatial distributions, the most
frequently adopted approach involves counting the number of sides of
each cell~\cite{Nishi_Hanasaki:1989, yacare:2004}.  Another
traditional approach which can be used to quantify polygonality
involves the use of the distance between nearest neighboring points as
well as derived measurement such as the conformity ratio
(e.g.~\cite{Wassle:1978, Cook:1996, Masland:2000}).  A comparison of
methods for characterization of spatial properties of retinal mosaics
has been presented by Cook~\cite{Cook:1996}, with emphasis on nearest
neighbor distances.

The present article starts by identifying the main properties required
from a good measurement of spatial order and follows by describing
each of the considered measurements, whose properties are then
evaluated for the characterization of global and local spatial order.
The analysis of the potential of the measurements to characterized
global properties takes into account simulated (hexagonal lattices
with progressive perturbations) as well as real data (retinal mosaics
of photoreceptors).  The possibility to use the measurements to
quantify local order around each point is also addressed with respect
to the identification of regions with distinct spatial order and the
analysis of a system involving progressive (radial) variation of
spatial order.

\section{Requirements}

Before considering possible indicators of spatial order and evaluating
their performance, it is important to identify the main features which
could be expected from a good such measurement. Such features are
discussed in the following.

\begin{figure}[h]
 \begin{center} 
   \includegraphics[scale=0.5,angle=0]{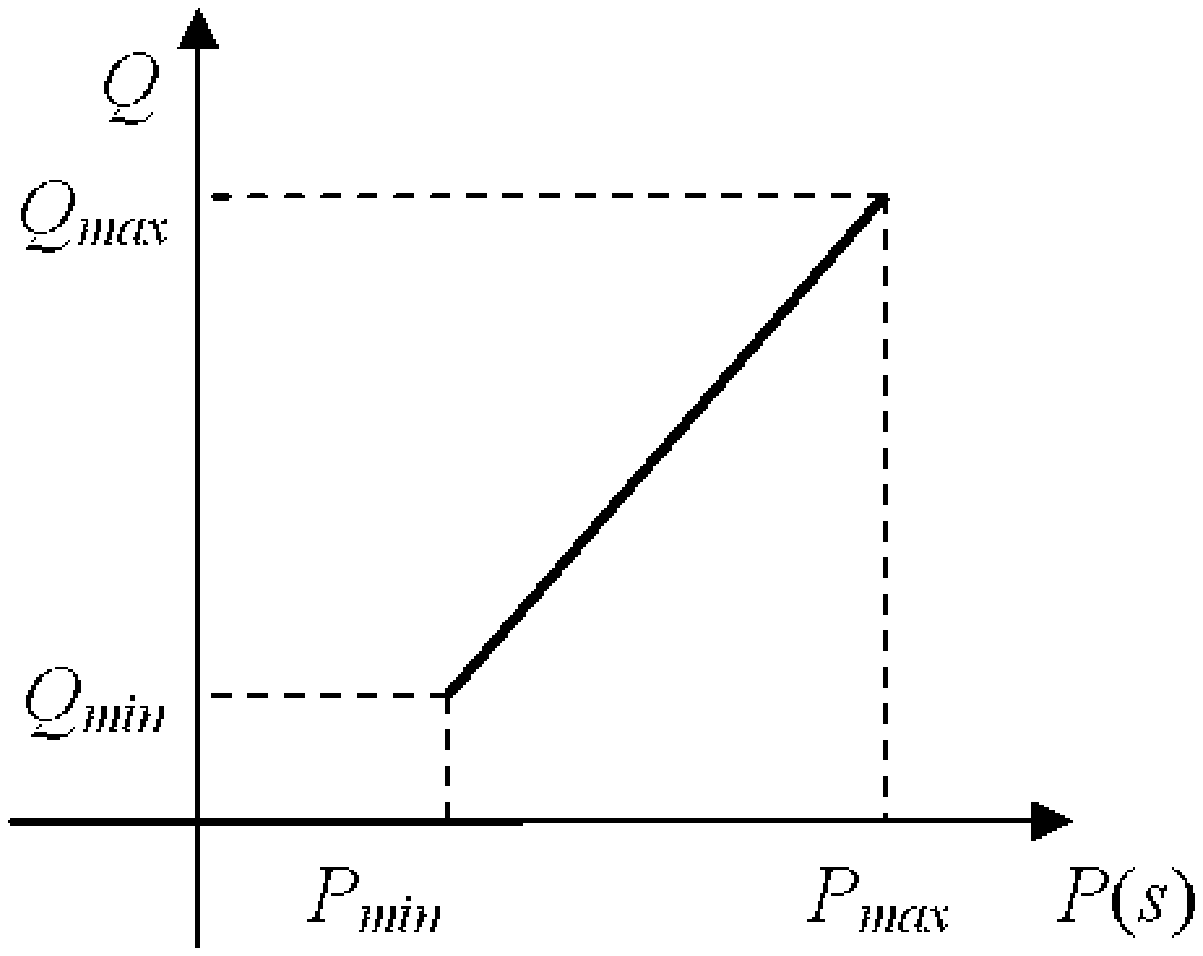}  \\
   (a) \\
   \vspace{0.3cm}
   \includegraphics[scale=0.5,angle=0]{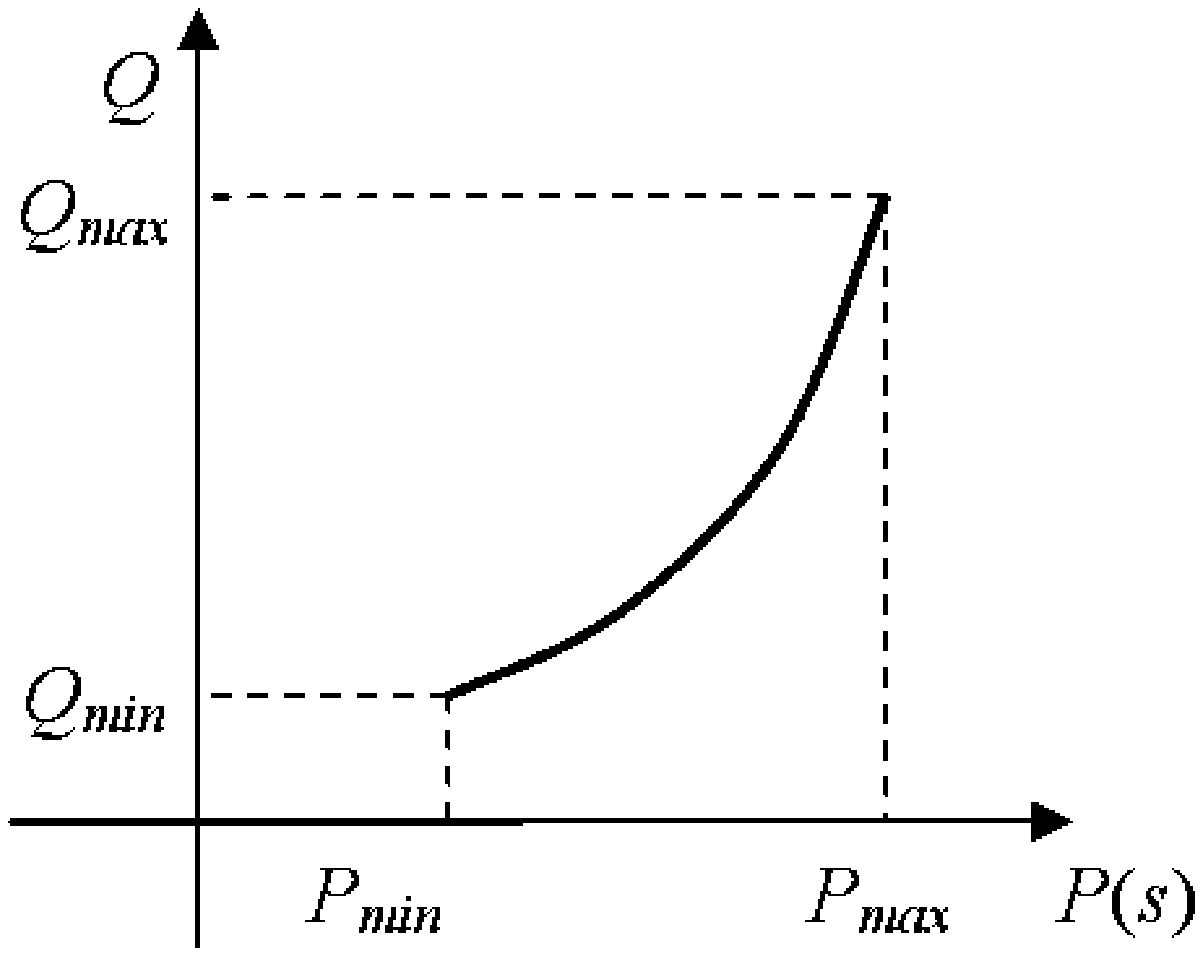}  \\
   (b) \\
   \vspace{0.5cm}
   \caption{Two possible measurements mapping the
    property $P$ into $Q$.~\label{fig:meas}}
\end{center}
\end{figure}

Let us consider that the property $P$ to be measured is a function of
a given parameter $s$~\footnote{The property $P$ may also be a
function of a series of parameters $\vec{s}$.  The situation
considered in this article assumes all but the single parameter $s$
are fixed.}, so that $P=P(s)$, and that the act of measuring $P$
involves mapping it into the quantity $Q(P(s))=Q(s)$.  For instance,
$s$ can be the intensity of spatial perturbation added to a perfectly
hexagonal lattice (see Section~\ref{sec:simul}), and $Q$ is the
measurement of spatial order.  Note that typically we do not have
access to the values of $P$, otherwise there would be no need to
consider its measurement.  Figure~\ref{fig:meas} illustrates mappings
from the property $P$ into two possible measurements $Q(s)$.  Although
this example assumes that $P$ is upper and lower bound, some
situations may imply unlimited values of $P_{min}$ or $P_{max}$.
Situations such as that depicted in Figure~\ref{fig:meas} can be
always normalized, by shifting the function along the $x$- and/or
$y$-axes, such that $\tilde{P}_{min}=\tilde{Q}_{min}=0$ and
$\tilde{P}_{max}=\tilde{Q}_{max}=1$, defining a linear relationship.
The mapping in (b) can also be linearized by using an additional
transformation of the curve itself.

\emph{Sensitivity/Linearity:} One important feature of any measurement
is its ability to respond to variations of the measured quantity.
Because measurements can be normalized, sensitivity should be
considered in relative terms along the interval of measurement of
interest.  For instance, the mapping in Figure~\ref{fig:meas}(a) is
more sensitive than that in (b) for small values of $P(s)$, but
becomes less sensitive for large values of $P(s)$.  The sensitivity of
the measurement in Figure~\ref{fig:meas}(a) is the same for any value
of $P(s)$.  Observe that the increase of sensitivity along one of the
regions of the mapping can only be achieved at the expense of loss of
sensitivity at another region.  Unless some specific interest of
achieving enhanced sensitivity at some specific range of values is to
be considered, measurements should ideally be linear.  In the specific
case of spatial order quantification, one is frequently interested in
the cases of small deviations from perfect order.  In such cases, it
is interesting to obtain greater sensitivity for higher values of
spatial order.

\emph{Generality, Uniformity and Invariance:} It is usually
interesting that the measurement of $P$ be monotonic (increasing or
decreasing), well-defined and with uniform properties along the whole
interval of measurements.  This implies absence of singularities
(e.g. tendency to go to infinity for some values of $P$) as well as
uniform sensitivity.  The linear mapping meets all such requirements.
In cases such as the quantification of spatial order of geometrical
systems, it is also important that the measurements be invariant to
translation, rotation and scaling.

\emph{Discriminative Power:} Although Figure~\ref{fig:meas} considered
a deterministic mapping from $P$ into $Q$, it often happens that
measurements taken over different realizations of a system for the
same parameter $s$ yield different values.  This is often the case
with spatial order.  For instance, the spatial order of the
photoreceptors in the retina of two similar animals (i.e. same age,
gender, etc.) from the same species will unavoidably be characterized
by distinct, even though similar, values.  In turn, such a stochastic
variation of $Q$ implies that in certain situations two realizations
of a system, respective to different parameters $s$, will yield the
same measurement value.  In other words, the stochastic variability of
$Q$ will tend to limit its \emph{discriminative power}.  Given two
instances of $s$, i.e. $s_1$ and $s_2$, the discriminative power will
vary inversely with the overlap between the statistical density
functions characterizing $P(s_1)$ and $P(s_2)$.  Such an overlap can
be estimated by considering the average and standard deviation of
those two densities, which is the procedure adopted in the present
article.

\emph{Adimensionality:} While this is not an imperative property, it
is interesting that the measurement in question be adimensional,
allowing it to be expressed in percentage values.  In addition,
adimensionality often facilitates multivariate statistical analysis
and testing (e.g.~\cite{Freund:1993,Hair:1998}).

\section{Methods}

\begin{figure}[h]
 \begin{center} 
   \includegraphics[scale=0.5,angle=0]{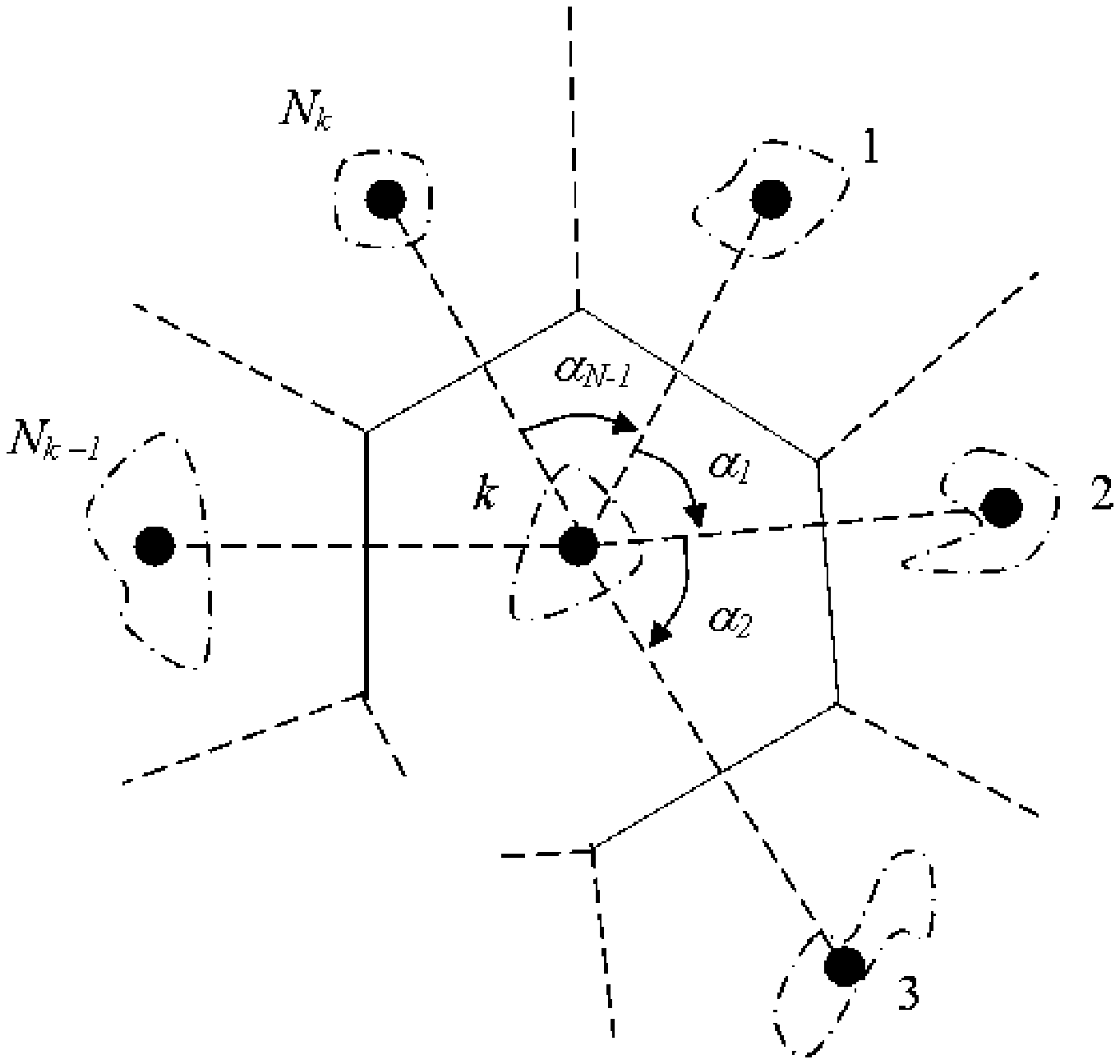}  \\
   (a)  \\
   \vspace{0.5cm}
   \includegraphics[scale=0.33,angle=0]{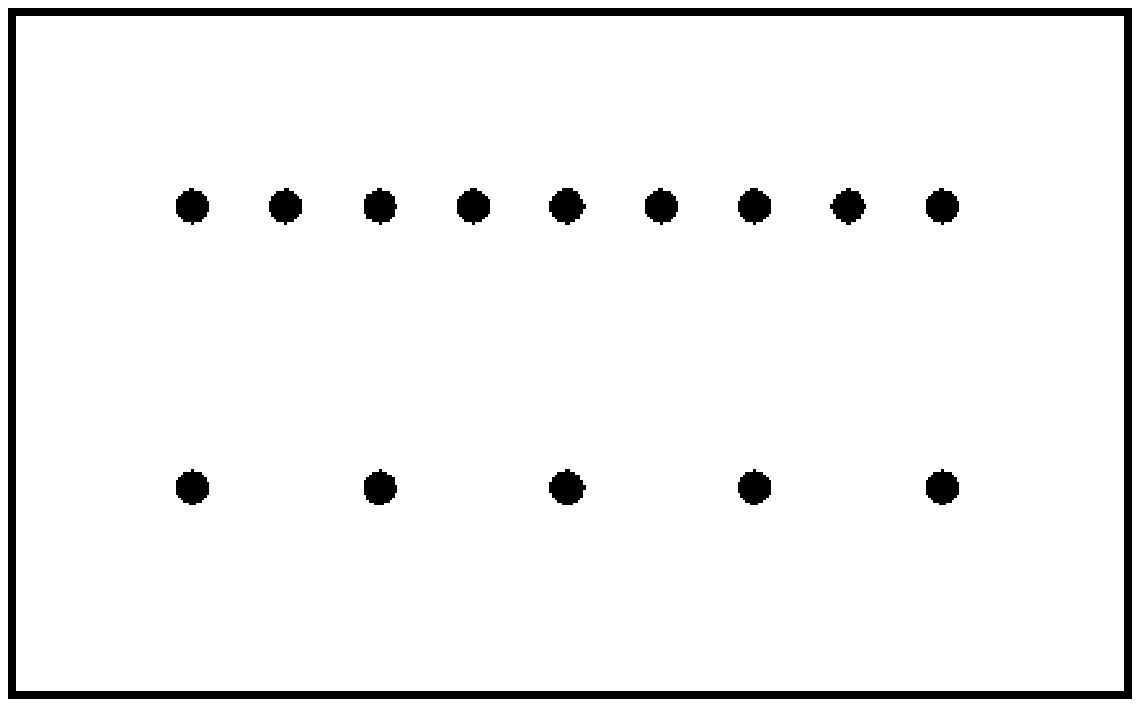}  
   \hspace{0.3cm}
   \includegraphics[scale=0.33,angle=0]{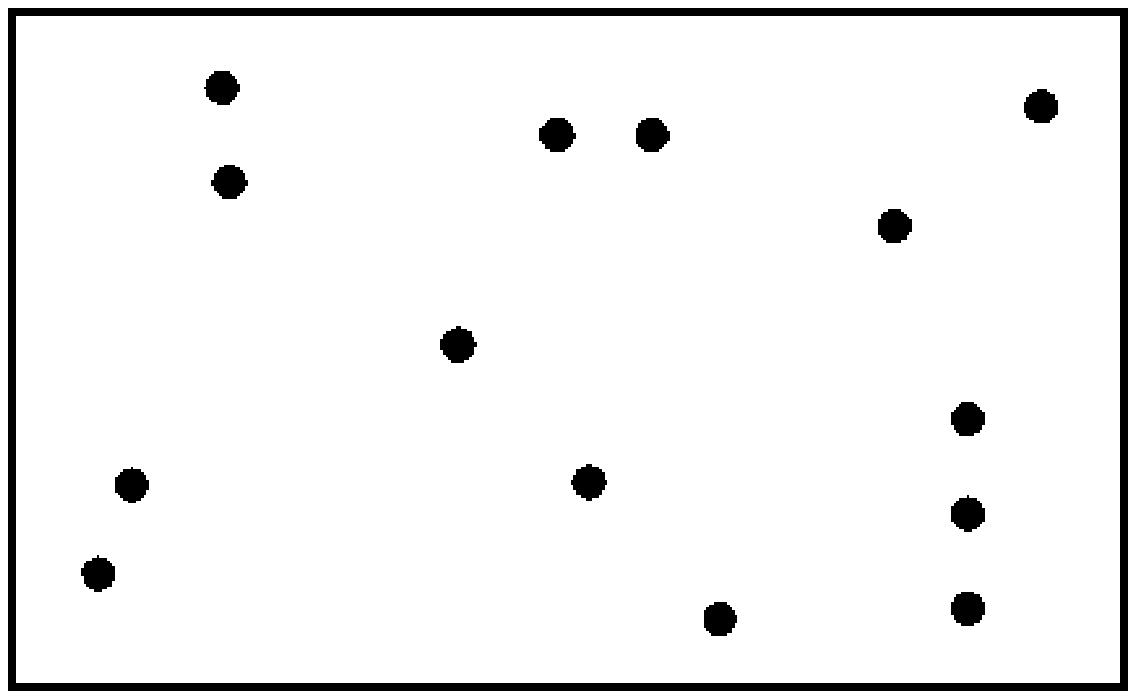}  \\
   (b) \hspace{3.5cm}  (c) 

   \caption{The geometrical construction and symbols used in the
   definition of the angular hexagonality are presented in (a).  The
   reference points are shown as black dots, and the neighbors of seed
   $k$ are enumerated in clockwise fashion from $1$ to $N_k$. Although
   rather distinct, the two point distributions in (b) and (c) yield
   identical statistics of shortest distances, and hence the same
   conformity ratio. ~\label{fig:equal}}

\end{center}
\end{figure}

Let the image containing the biological system under analysis be
represented in an $N_i \times N_j$ image $A$.  Each element of the
system is identified by an integer label $i$, while its spatial
position is indicated in terms of the coordinates $(x(i),y(i))$ of
some \emph{reference point} or \emph{seed}, often corresponding to the
centroid of the elements (see Figure~\ref{fig:equal}a).  The
\emph{nearest neighbor distance} between a point $i$ and a set of
points $S$ is defined as the smallest distance between $i$ and each of
the points in $S$.  The Voronoi tessellation of the system of
reference points can be obtained by assigning to each pixel $A(i,j)$
the label of the nearest reference point.  One of the nice features of
Voronoi tessellations is that they can be used to establish boundaries
and adjacencies between the original reference points.  More
specifically, given a reference point $k$, each of its Voronoi
neighbors will correspond to a reference point $i$ such that the
Voronoi cells of $k$ and $i$ share a common side. In order to avoid
border effects, the Voronoi cells which are adjacent to the image
background (i.e. the border cells) are henceforth identified and
excluded from the calculations.

{\bf Number of Voronoi cells:} An interesting measurement which can be
immediately obtained from Voronoi tessellations corresponds to the
number of sides of each Voronoi cell, a measurement which has been
called \emph{hexagonality} (e.g.~\cite{Nishi_Hanasaki:1989,
yacare:2004}).  Henceforth $n_k$ represents the number of sides of the
cell associated to the reference point $k$.  An hexagonal system, for
example, will produce all cells with 6 sides, except for the cells at
the border of the system.

\begin{figure}[h]
 \begin{center} 
   \includegraphics[scale=0.6,angle=0]{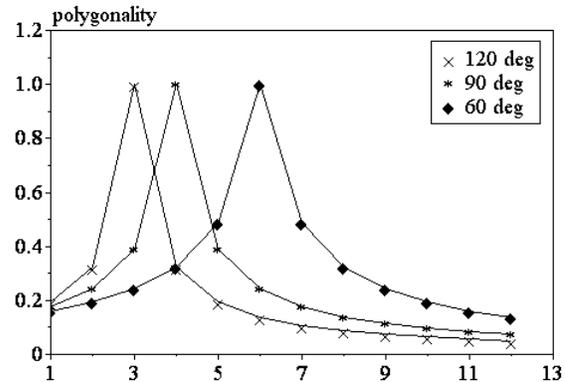}  \\

   \caption{The poligonalities for $\beta = 120$, $90$ and $60$ degrees
   obtained for a single point surrounded by $N$ uniformly distributed
   angles (i.e. $\alpha = 2 \pi/N$).  Observe that the sharp peaks
   clearly identify each type of symmetry.  For instance, the perfect
   hexagonal arrangement of neighbors for $N=6$ implies the sharp peak
   of the poligonality assuming $\beta=60$ degrees at that respective
   position.  ~\label{fig:polygs}}

\end{center}
\end{figure}

{\bf Angular Regularity:} For each Voronoi cell $k$, starting at an
arbitrary neighbor $i$, determine the successive $N_k$ neighbors as
one turns around the cell in a clockwise fashion and calculates the
angles $\alpha_i$ defined by the adjacencies (see
Figure~\ref{fig:equal}a).  The angular polygonality of a given point
$k$ can then be expressed in terms of the sum of angle differences for
each point, i.e.

\begin{equation}  \label{eq:sum_ang}
  \Sigma= \sum_{i=1}^{N_k} | \alpha_i - \beta |
\end{equation}  

where $\beta$ is a specific angle of interest. In case one is
interested to quantify the \emph{angular hexagonality} of a point
system, $\beta$ should be taken as 60 degrees.  Similarly, in case one
wants to check for orthogonal arrangement, $\beta$ should be 90
degrees.  Other values of $\beta$ varying in the interval $0 < \beta <
180^o$ can also be used in order to check for other types of angular
distributions.  However, only 60, 90 and 120 degrees allow full
regular tiling of the plan.  Figure~\ref{fig:polygs} shows the
polygonalities of a single point surrounded by $N$ points (uniform
angle spacing).  Note that the sum of angle differences varies from 0
to infinite (i.e. a point surround by an infinite number of points).
Although it would be possible to consider the average of angle
differences (i.e. divide the sum of angle differences in
Equation~\ref{eq:sum_ang} by the number of neighbors of each point),
such an alternative would lead to similar measurements as the number
of neighboring cells increase, therefore reducing the discriminative
power.

An alternative measurement of spatial order, yielding the adimensional
measurement henceforth called the \emph{polygonality index}, can be
defined as a function of the sum of difference angle as follows

\begin{equation}  \label{eq:hex}
  \Delta_{\alpha}(k) = \frac{1}{\sum_{i=1}^{N_k} | \alpha_i - \beta |+1}
\end{equation}  

This measurement, which was first considered for hexagonality
characterization in~\cite{hexa:2005}, is bound between 0 (lack of
spatial order) to 1 (perfect polygonality).

{\bf Conformity Ratios:} Once all nearest neighbor distances have been
calculated, for instance by using the simple and effective algorithm
described in~\cite{Cook:1996}, the conformity ratio of distances can
be defined~\cite{Wassle:1978} as the ratio between the mean and
standard deviation of all nearest neighbor distances.  More organized
systems should therefore imply higher conformity ratios.

{\bf Coefficient of Variation:} The exact inverse of the conformity
ratio, the coefficient of variation is also considered in this work
because of its traditional use in statistics.

Note that the conformity ratio of the nearest distances present a
singularity (tends to infinity) when the shortest distances for every
point in the system is equal, as with regular lattices.  The
coefficient of variation does not present singularities, but will
produce null values for any regular lattice (null standard deviation
of shortest distance will imply coefficient of variation equal to
zero).

Although it would be possible to define conformity ratio and
coefficient of variation of ~\emph{angles differences}, such
measurements would be completely unable to cope with situations such
as that considered in Figure~\ref{fig:polygs}, i.e. involving $N$
equal (or similar) angles around a point.  In these cases, the
standard deviation of the angle values is null, implying conformity
ratio equal to zero and coefficient of variation equal to infinity.
Such a property also implies lack of sensitivity for systems
characterized by a high level of spatial order. Contrariwise, the sum
of angle differences or hexagonality index will produce distinct
values for any number of equally spaced neighboring points.  For such
reasons, and also because of poor performance in preliminary
experimental investigations conducted by the authors, the coefficient
of variation and conformity ratio of angle differences will not be
considered further in this article.

Although all the above measurements are invariant to translation,
rotation and scaling of the system under analysis, \emph{any}
measurement derived from the minimum distance statistics will fail to
distinguish between situations such as those illustrated in
Figure~\ref{fig:equal}(b) and (c), which yield identical statistical
densities of shortest distances and therefore equal conformity ratios.
This important limitation of the shortest distance measurement stems
from the fact that it only takes into account the most immediate
neighborhood around each point, overlooking spatial properties at
higher spatial scales.  Therefore, a natural means to try to obtain
better discriminative power while using distance statistics consists
in considering successive shortest distances, such as the second,
third, etc., shortest distances.  Such a possibility is also explored
in this work.

A severe limitation of the method of counting the number of Voronoi
cell sides follows from the fact that the number of cell sides will
not change except for gross perturbations of the particle system.
However, this measurement can be potentially useful when the
structures under analysis involve large perturbations.

\section{Results}

This section describes a systematic performance comparison of methods
used to quantify the spatial order of point distributions, including
nearest distances, number of Voronoi sides, conformity ratio and
coefficient of variation of the nearest distance, sum of angle
differences and hexagonality index.  This evaluation considers the
quantification of \emph{global} and \emph{local} spatial order in
systems of points.  While the former case addresses the problem of
assigning a single measurement to the whole system of points, the
latter situation involves the quantification of the spatial order
around each point of the system.  The global evaluation is performed
for simulated data (regular hexagonal lattices with progressive
spatial perturbations) and real data concerning agouti
(\emph{Dasyprocta agout}) retinal photoreceptor mosaics.  The obtained
results clearly corroborate the superiority of the sum of difference
angles and the hexagonality index, especially regarding their
linearity, sensitivity and discriminative power.  The investigation of
the potential of the considered measurements for quantifying local
spatial order involves the identification of regions with different
levels of order as well as a system where the spatial order varies
radialy.

\subsection{Simulated Data}~\label{sec:simul}

\begin{figure*}[h]
 \begin{center} 
   \includegraphics[scale=0.55,angle=0]{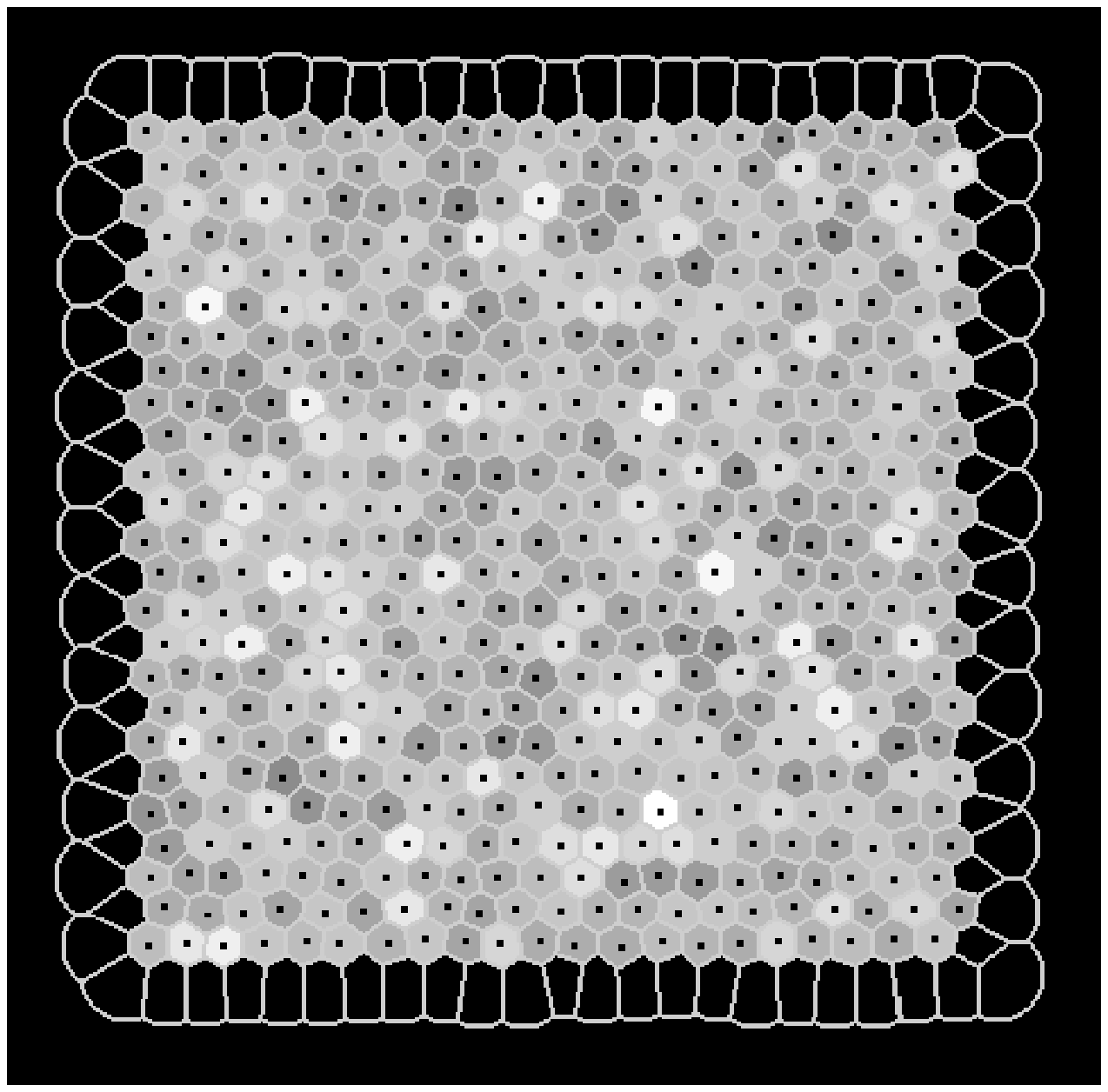} 
   \hspace{1.5cm}
   \includegraphics[scale=0.55,angle=0]{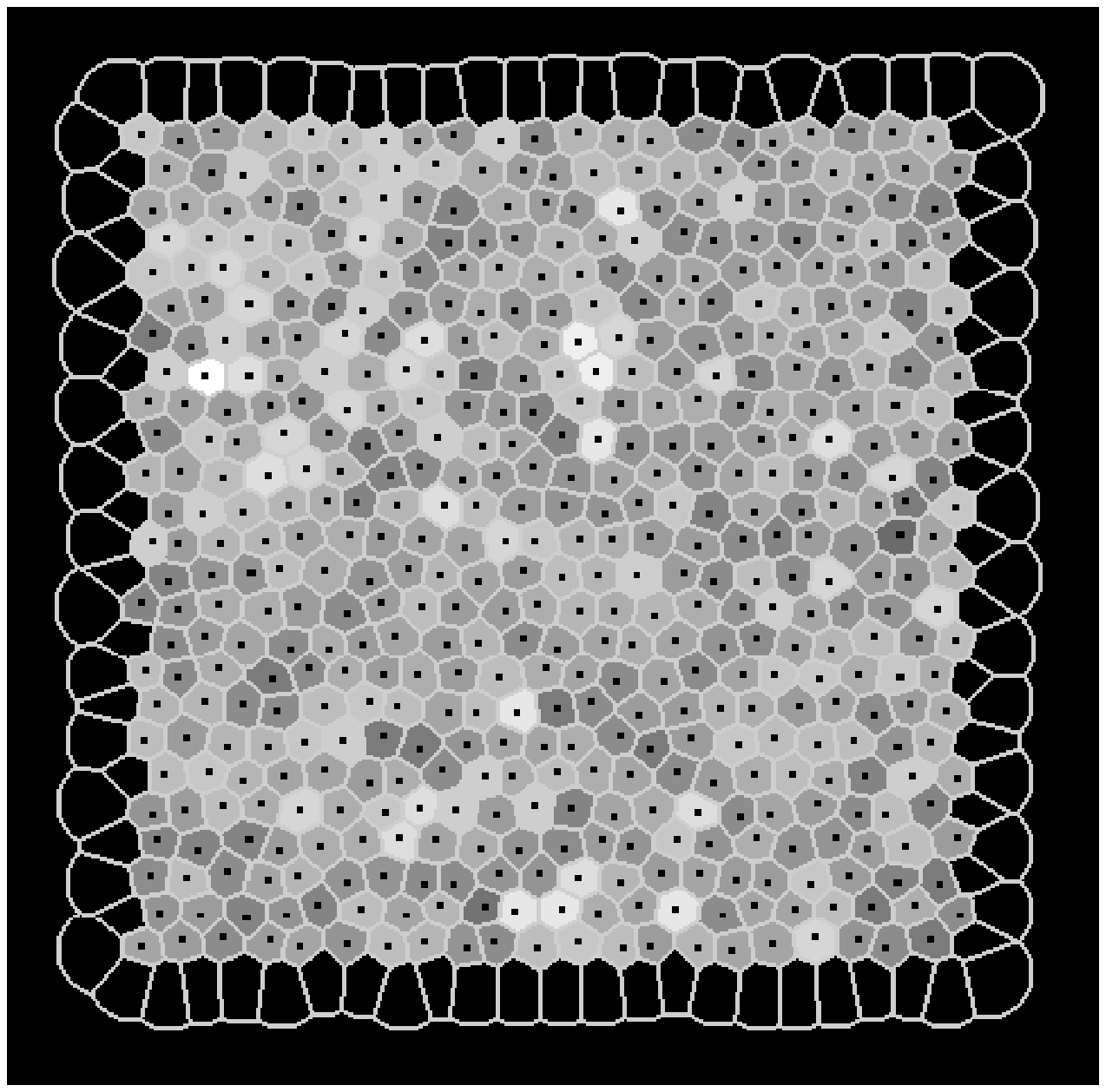} \\
   (a)   \hspace{8.5cm}  (b) \\
   \vspace{0.5cm}

   \caption{Examples of simulated hexagonal lattices with
   perturbations $\delta=2$ (a) and $4$ (b).  The hexagonality index
   is expressed by the gray levels at each Voronoi cell, with higher
   values indicated in light gray.  Note the border elements in black,
   whose hexagonality is not considered. The minimum and maximum gray
   levels in each image are normalized between black and white for the
   sake of better visualization.~\label{fig:hex}}

\end{center}
\end{figure*}

\begin{figure*}[h]
 \begin{center} 
   \includegraphics[scale=0.65,angle=0]{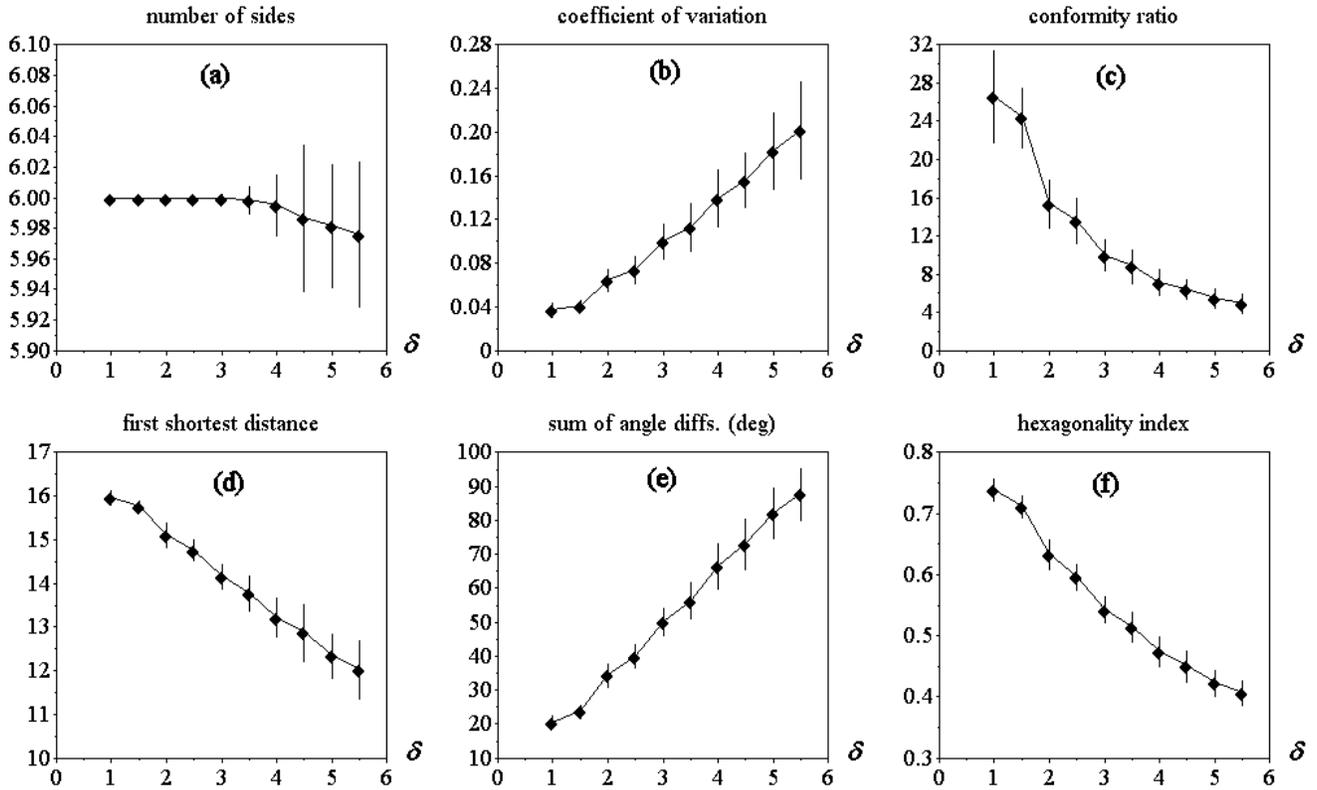}  \\
   \vspace{0.5cm}

   \caption{The average $\pm$ standard deviation of the number of
   Voronoi cell sides (a), coefficient of variation of first shortest
   distance (b), conformity ratio of first shortest distance (c) first
   shortest distances (d), average of angle differences (e), and
   hexagonality indices (f), in terms of the perturbation intensity
   $\delta$ considering all simulations. For the sake of better
   visualization, the standard deviations are shown to 5 times their
   original values.~\label{fig:graphs}}

\end{center}
\end{figure*}

In order to investigate the performance of the above measurements
under a controlled situation, hexagonal lattices with progressive
spatial perturbation intensities were obtained by using
mathematic-computational means.  More especifically, hexagonal lattice
with 621 points and vertices of length $\Delta = 10$ pixels were
generated in a $480 \times 480$ pixels image and progressively
perturbated with uniformly distributed displacements of magnitude
$\delta = 1, 2, \ldots, 10$. A total of 50 realizations of each
configuration (i.e. each perturbation intensity) was performed in
order to enhance statistical representativity.  Figure~\ref{fig:hex}
shows two of the considered simulated hexagonal lattices, respective
Voronoi tessellations and hexagonality indices obtained for
perturbation intensities $\delta=2$ and 4, respectively.  The
histograms of the number of Voronoi cell sides, coefficient of
variation of shortest distances, conformity ratio of shortest
distances, shortest distance values, average angle difference (in
degrees), and hexagonality indices for each of the perturbation
intensities are shown in Figure~\ref{fig:graphs}.

\begin{figure*}[h]
 \begin{center} 
   \includegraphics[scale=0.6,angle=0]{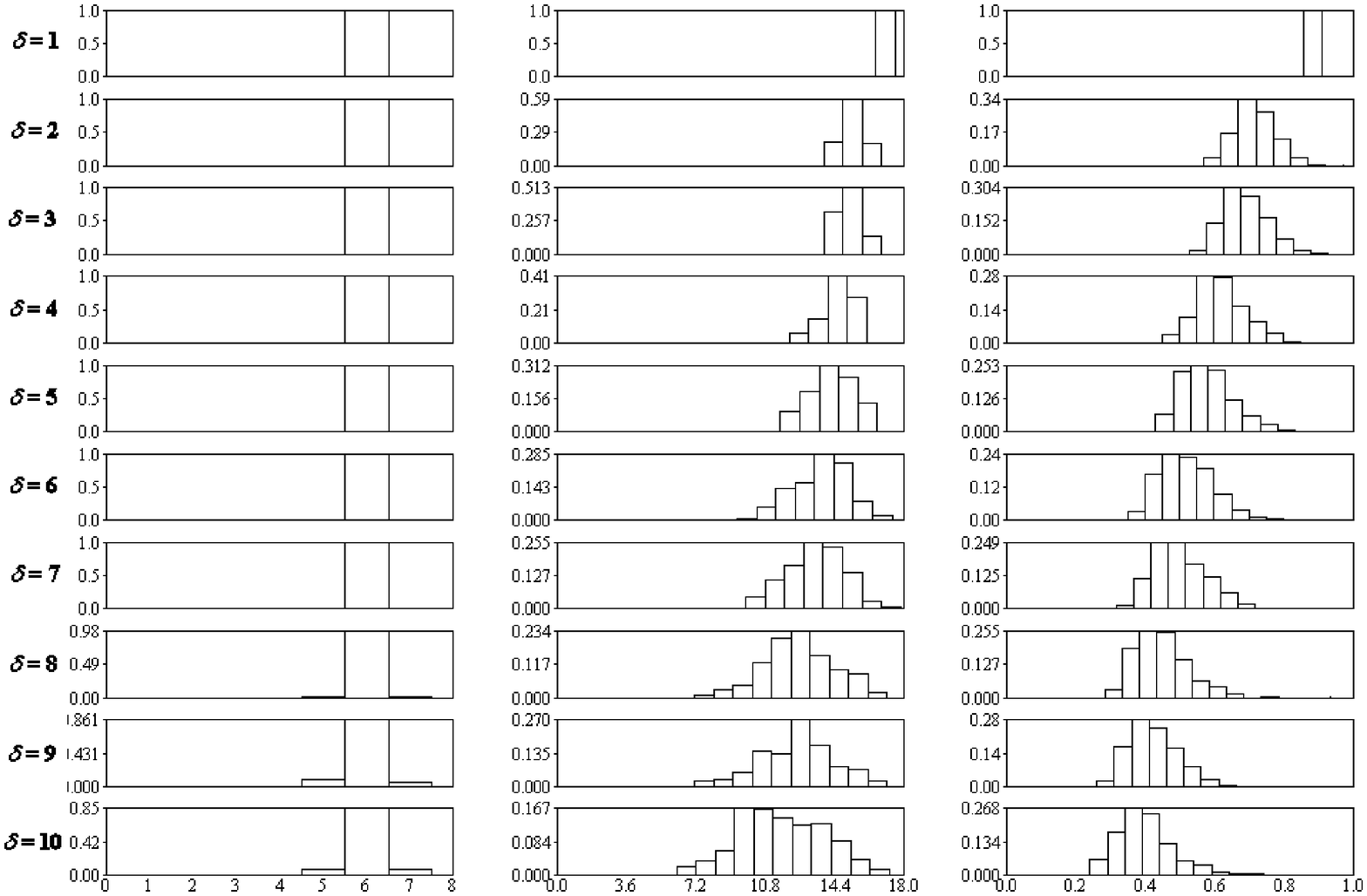}  \\

   \caption{The histograms of number of Voronoi cell sides (first
   column), nearest neighbor distances (second column) and
   hexagonality indices (third column) obtained for 50 realizations of
   each of the configuration, with progressive perturbation
   intensities $\delta$.~\label{fig:avst}}

\end{center}
\end{figure*}

It is clear from these results that the number of Voronoi cell sides,
shown in Fig.~\ref{fig:avst}(a), is largely invariant to the
perturbation intensity.  In the cases where the latter is large enough
to induce changes, it acts mainly by increasing the dispersion of the
measurements while keeping an almost constant average value.  The
coefficient of variation (b), first shortest distance (d) and sum of
angle differences (e) all resulted in an almost linear mapping between
the perturbation intensity and the measurement. All these cases, and
especially the coefficient of variation, are also characterized by an
increase of dispersion for larger perturbations.  The conformity ratio
resulted the most non-linear relationship, providing greater
sensitivity for smaller perturbations.  However, the measurement
dispersions are also substantially larger in such cases, which limits
the discriminative power of this measurement.  The hexagonality index,
shown in Fig.~\ref{fig:avst}(f), accounts for an almost linear
mapping, with a slight increased sensitivity for small perturbations.
In addition to these interesting features, this measurement is the
only one characterized by an almost constant standard deviation which,
combined with the almost linear mapping characterizing this
measurement, implies a constant discriminative power along all the
considered perturbation intensities.  The linearity, sensitivity and
discriminative power of the measurements of spatial order are
summarized in Table~\ref{tab:meas}.  It follows from such results
that the hexagonality index is particularly suitable in cases of
higher spatial order, while the sum of angles provides an interesting
alternative for analysis of less ordered systems.

\begin{table*}
  \vspace{1cm}
  \begin{tabular}{||l|l|l|l||}  \hline
   \bf{Measurement}    &   \bf{linearity} & \bf{sensitivity} & \bf{discriminative power} \\ \hline
   \bf{Number of sides}  &  very smal  &    very low &    very low \\
   \bf{Coef. of variation}  &    high     & almost constant &  higher for small perturbations \\
   \bf{Conformity ratio}    &   small     & higher for small perturbations &  higher for large perturbations \\
   \bf{shortest distance}   &   high      & almost constant &  higher for large perturbations \\
   \bf{sum of angle diffs.} &   high      &  almost constant & higher for small perturbations \\
   \bf{hexagonality index}  &   medium    & slightly higher for small perturbations &  high and almost constant \\  \hline
  \end{tabular}
\caption{The linearity, sensitivity and discriminative power of the
considered measurements of spatial order.}~\label{tab:meas}
\end{table*}

The properties of the number of sides, shortest distance and
hexagonality index are confirmed by analysis of the histograms in
Figure~\ref{fig:avst}, which show the distribution of such
measurements for several perturbation intensities.

\subsection{Biological Data}~\label{sec:agouti}

In order to further evaluate the performance of the conformity ratio
and sum of angle differences, these two measurements have been applied
to characterize the regularity of spatial distribution of two types of
cones in agouti photoreceptor mosaics.  The choice of the sum of
angles instead of the hexagonality index was motivated by the fact
that such systems involve relatively high levels of disorder.

The agouti (\emph{Dasyprocta agout}), is a hystricomorph rodent whose
diurnal habit and well-developed visual streak yield an interesting
model for comparative studies of the visual system
(e.g.~\cite{Silveira:1989, Picanco:1991}). The organization of mosaics
formed by short-wavelength-sensitive cones (S-cones) and middle- to
long-wavelength-sensitive cones (M/L-cones) photoreceptors in the
agouti's retina have been analyzed. For this purpose we used two
polyclonal antibodies that have been shown to label S-cones (JH455) or
M/L-cones (JH492) in a range of mammals, using immunocytochemical
methods largely according to the procedures described
elsewhere~\cite{Ahnelt}. Using $40\times$ oil immersion objective, 25
fields taken with $250 \times 250 \mu m$ for M/L cones and 23 fields
with $500 \times 500 \mu m$ for S cones were acquired along the
vertical-dorsal axis of the retina with the aid of an optical
microscope (Eclipse E600, Nikon, Japan) equipped with a
high-resolution video camera (Nikon 4500). The raw images of the cones
were captured at the level of the inner segments. For carrying out the
analysis, $x$ and $y$ coordinates were identified by using the Scion
Image Software (ScionCorp).

\begin{figure}[h]
 \begin{center} 
   \includegraphics[scale=0.52,angle=0]{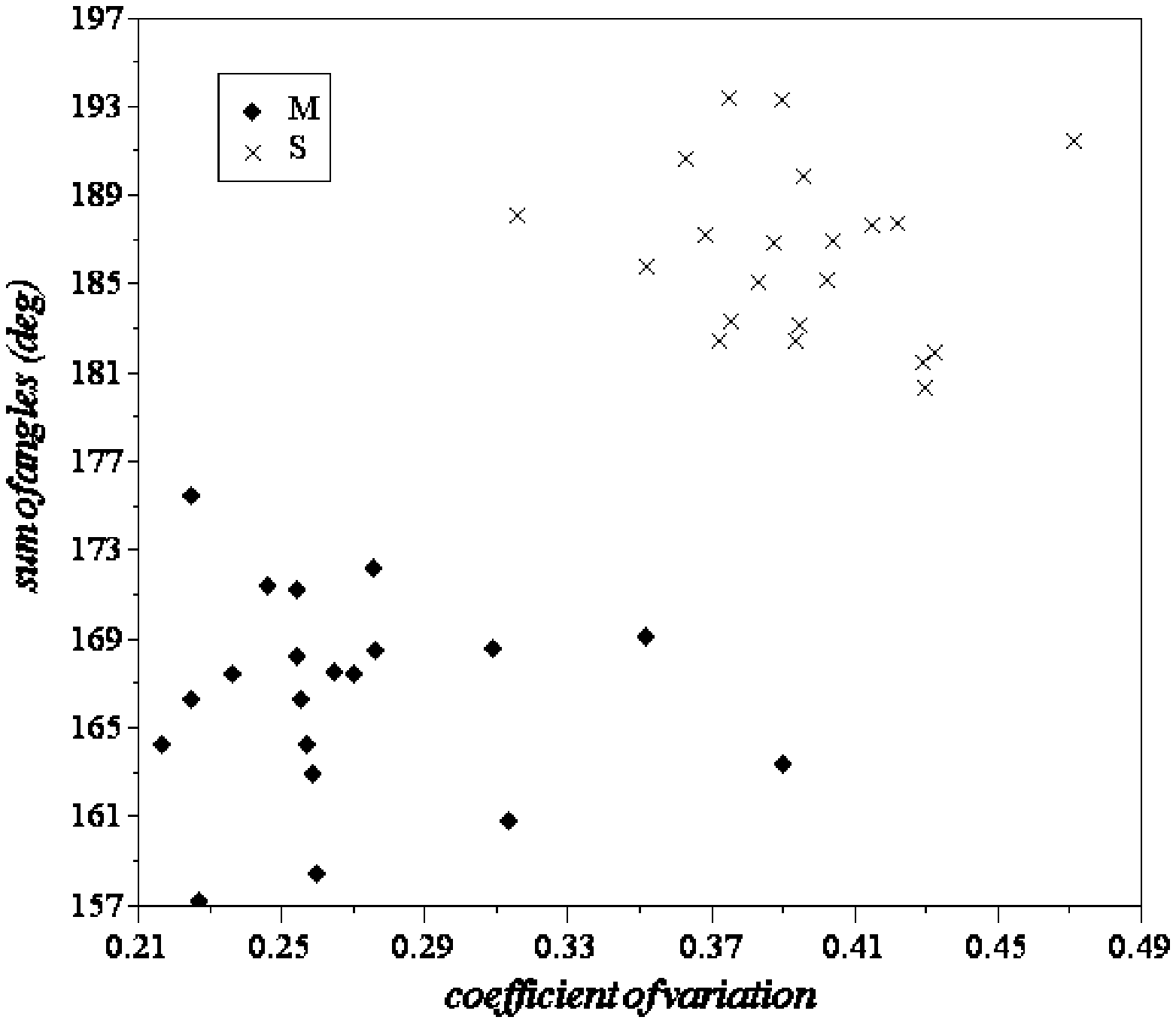}  \\

   \caption{Average conformity ratio and sum of angle differences
   obtained for each of the cone mosaics.  Note that the $S$ and $M/L$
   cone populations can be completely separated by considering the sum
   of angle differences isolately, which is not possible in the case
   of the conformity ratio. On the other hand, the conformity ratio
   allowed the identification of a high variability of the $M/L$ cone
   arrangements.~\label{fig:scatter}}

\end{center}
\end{figure}

Figure~\ref{fig:scatter} shows the average conformity ratios
($x$-axis) and sum of angle differences ($y$-axis) obtained for each
of the considered photoreceptor areas.  It is clear from this figure
that the two cone populations are clearly separated by considering the
sum of angle differences, while substantial overlap is observed for
the conformity ratio.  The obtained separation is even more definite
than previous results obtained for the same data set by using the
lacunarity measurement of translational
invariance~\cite{hexa_lacun:2005}.  The fact that the conformity ratio
allowed the characterization of higher variability of the $M$ cones
suggests that additional insights can be obtained by using
combinations of these two measurements.

In order to investigate the effect of using successive shortest
distances from each point, conformity ratios and coefficient of
variations were obtained considering the second to the fifth shortest
distances.  Discriminant analysis~\cite{Hair:1998,Duda_Hart:2001}
~\footnote{This type of statistical analysis involves the
identification of the project from a high dimensional measurement
space into a few dimensions so as to optimize the separation between
the involved classes of objects.} was then applied in order to
identify the contribution of each of such measurements to the overll
separation between the two types of mosaics.  The results indicated
that the sum of angles is the only measurement among all possible
linear combination of the considered features which allows all the
mosaic types to be correctly identified.  The use of conformity ratios
and coefficients of variation for several shortest distances always
implied in misclassifications of the mosaics.

\subsection{Characterization of Local Spatial Order}

\begin{figure*}[h]
 \begin{center} 
   \includegraphics[scale=0.7,angle=0]{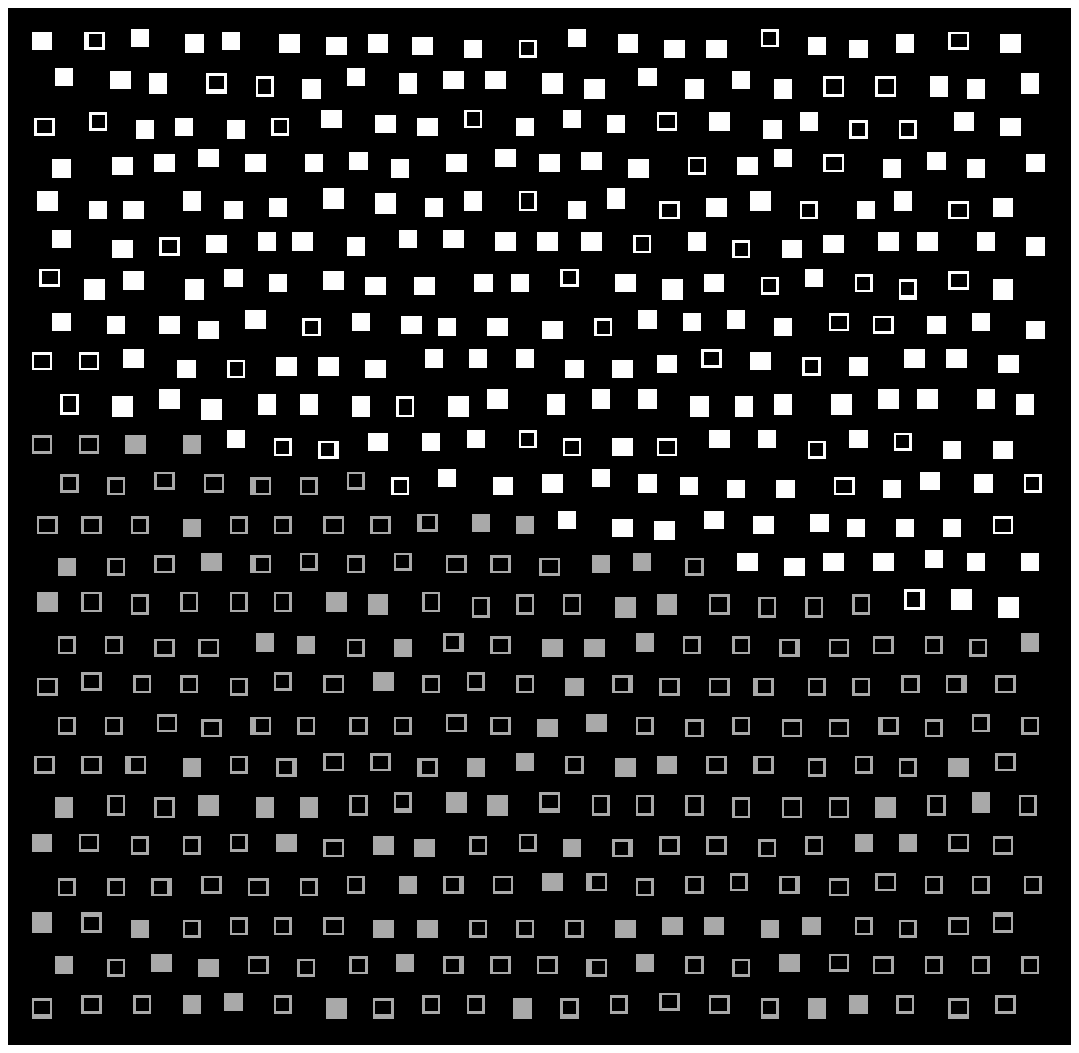} 
   \hspace{1.5cm}
   \includegraphics[scale=0.7,angle=0]{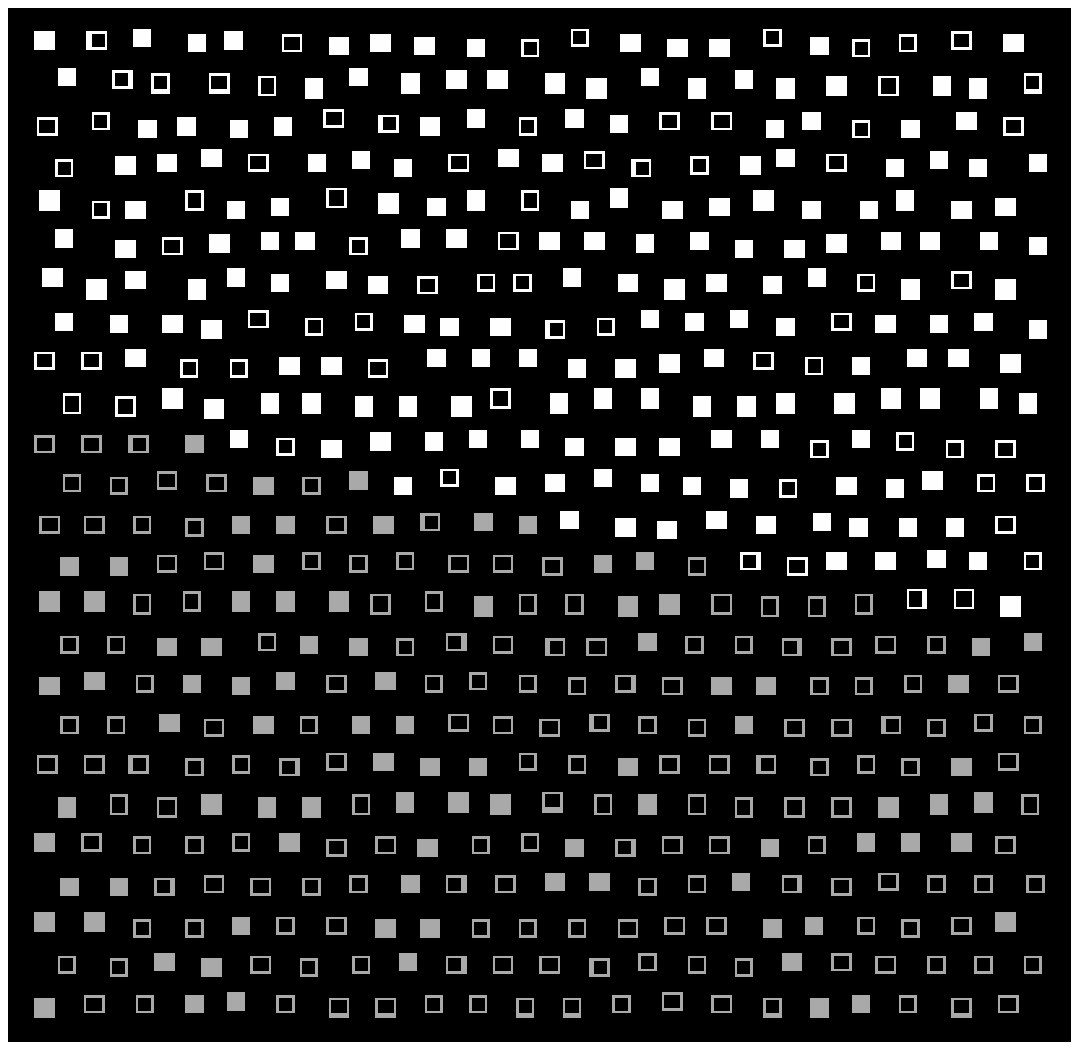}  \\
   (a)  \hspace{9cm}  (b) \\
   \vspace{0.3cm}
   \includegraphics[scale=0.7,angle=0]{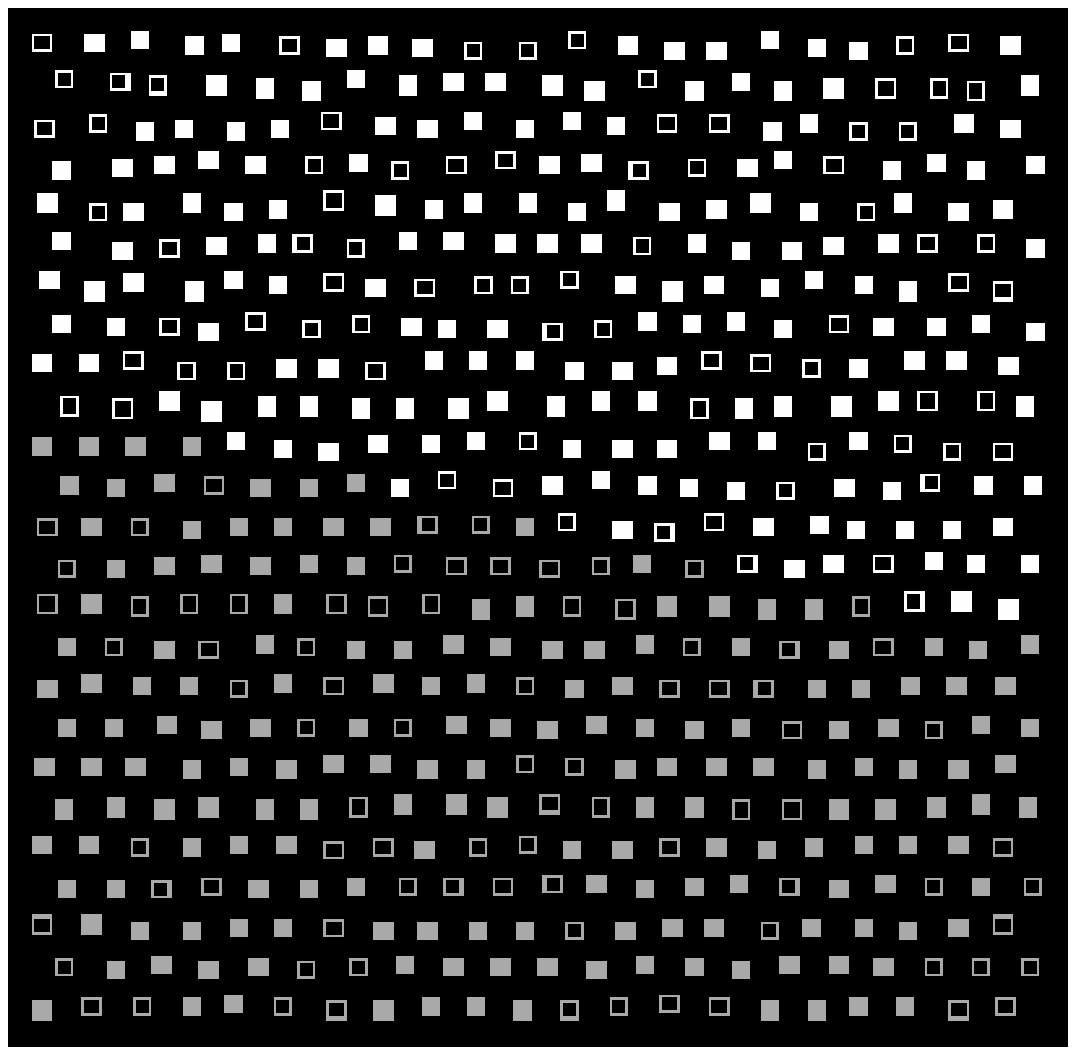} 
   \hspace{1.5cm}
   \includegraphics[scale=0.7,angle=0]{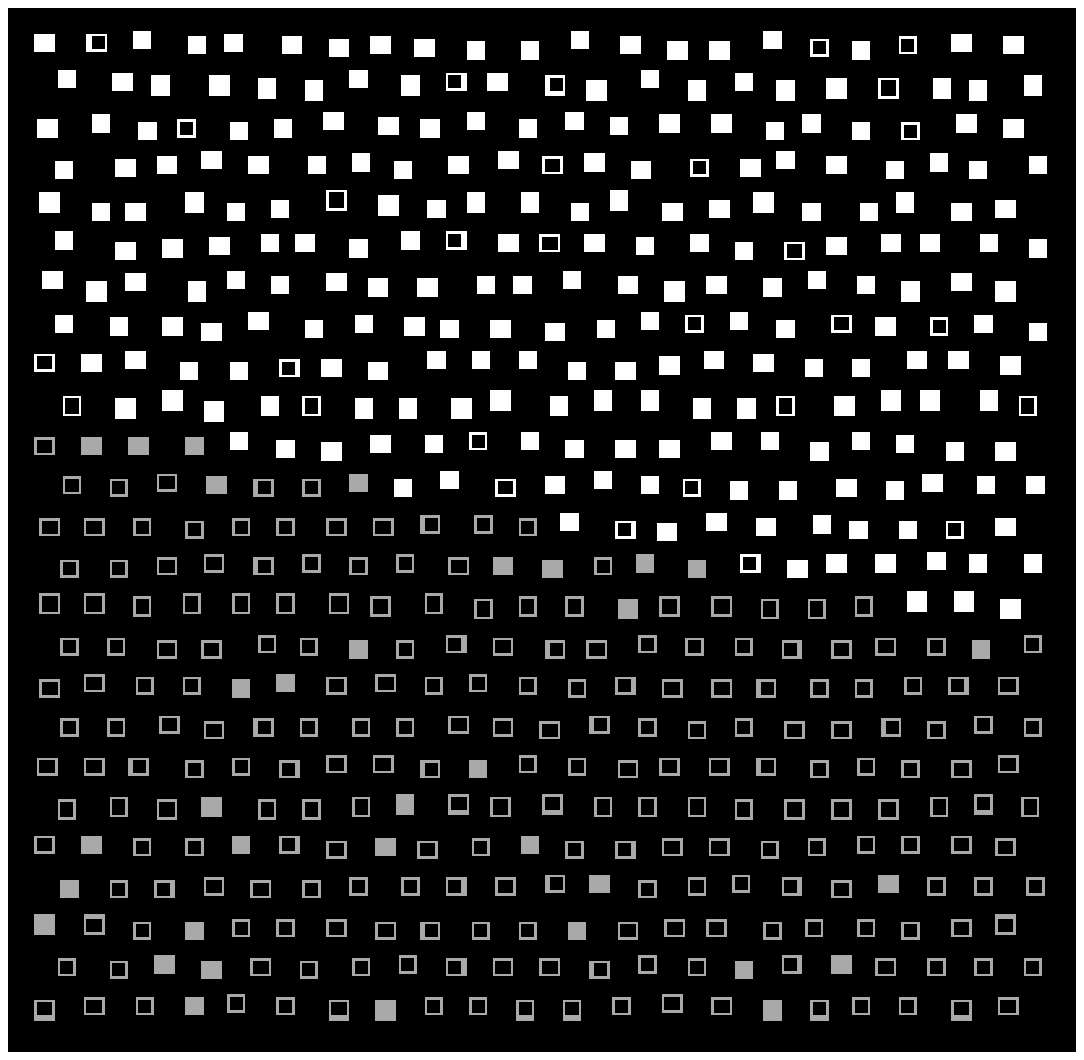} \\
   (c)  \hspace{9cm}  (d) \\
   \vspace{0.5cm}

   \caption{Local order identification of a system of points involving
   two regions with distinct spatial order by using the first (a),
   second (b) and third (c) shortest distances and the hexagonality
   index (d). The squares with holes correspond to the points whose
   measurements were more likely to belong to the lower domain (higher
   spatial order.)~\label{fig:local_dom}}

\end{center}
\end{figure*}

Although we have so far concentrated on the characterization of the
global properties of the spatial order in systems of points, another
relevant problem involves the quantification of \emph{local} order
around each point.  This issue is particularly important because,
provided one can quantify the spatial order around a small
neighborhood of each point, it becomes possible to study how such a
measurement varies along the whole system of points.  A particularly
interesting example of such cases, which often appears in complex
systems, is the existence of multiple \emph{spatial domains}
characterized by varying levels of spatial order.  This situation is
illustrated in Figure~\ref{fig:local_dom}, including two simulated
regions with different levels of spatial order.  These domains,
separated by an oblique border at the middle of the image, were
obtained by perturbing a hexagonal lattice with $\delta=2$ and $4$,
respectively.  Such a relatively small variation of spatial order was
intentionally imposed in order to impose a more demanding
discrimination task on the measurements.

\begin{figure}[h]
 \begin{center} 
   \includegraphics[scale=0.6,angle=0]{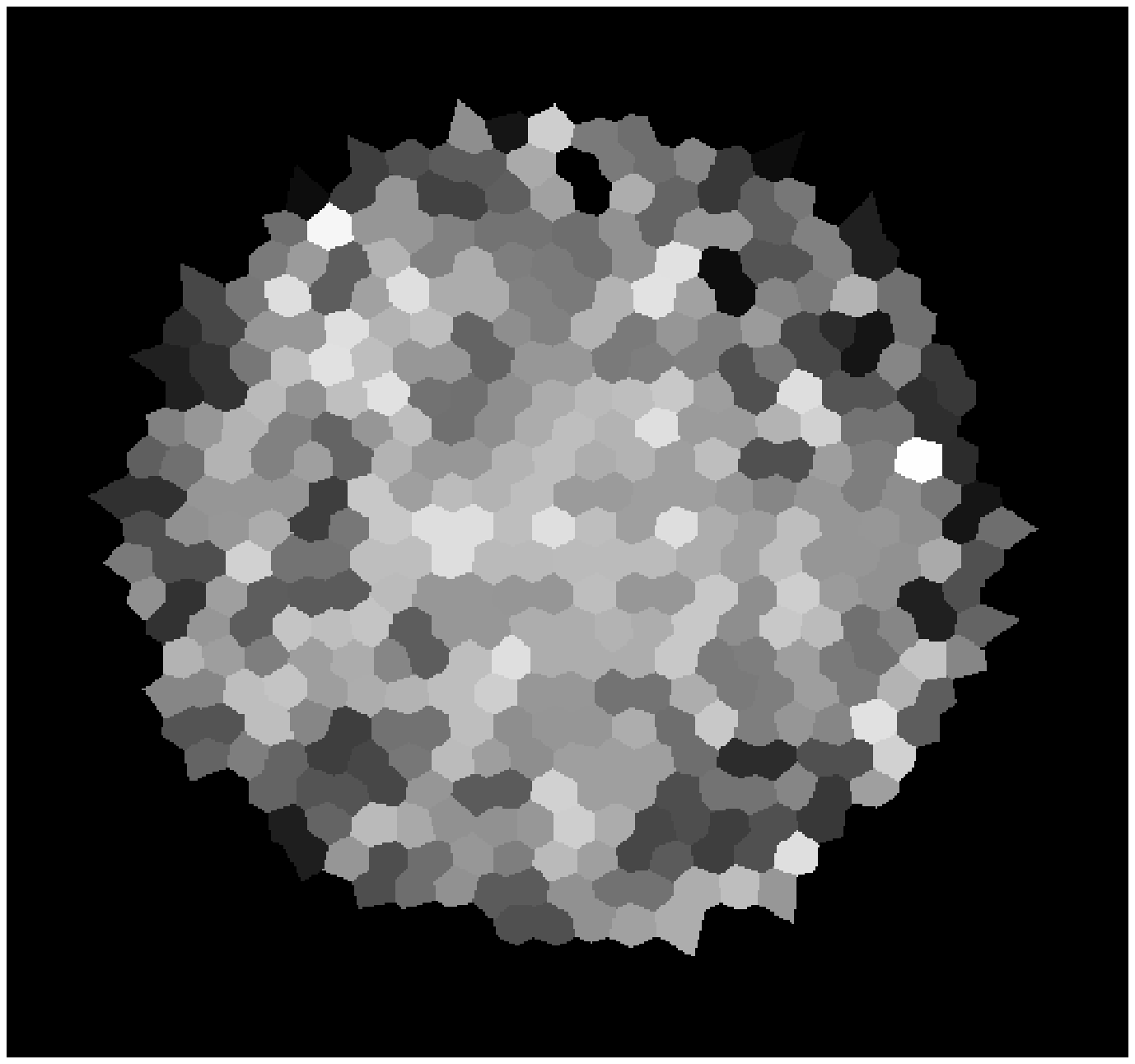} \\
   (a)  \\
   \vspace{0.3cm}
   \includegraphics[scale=0.6,angle=0]{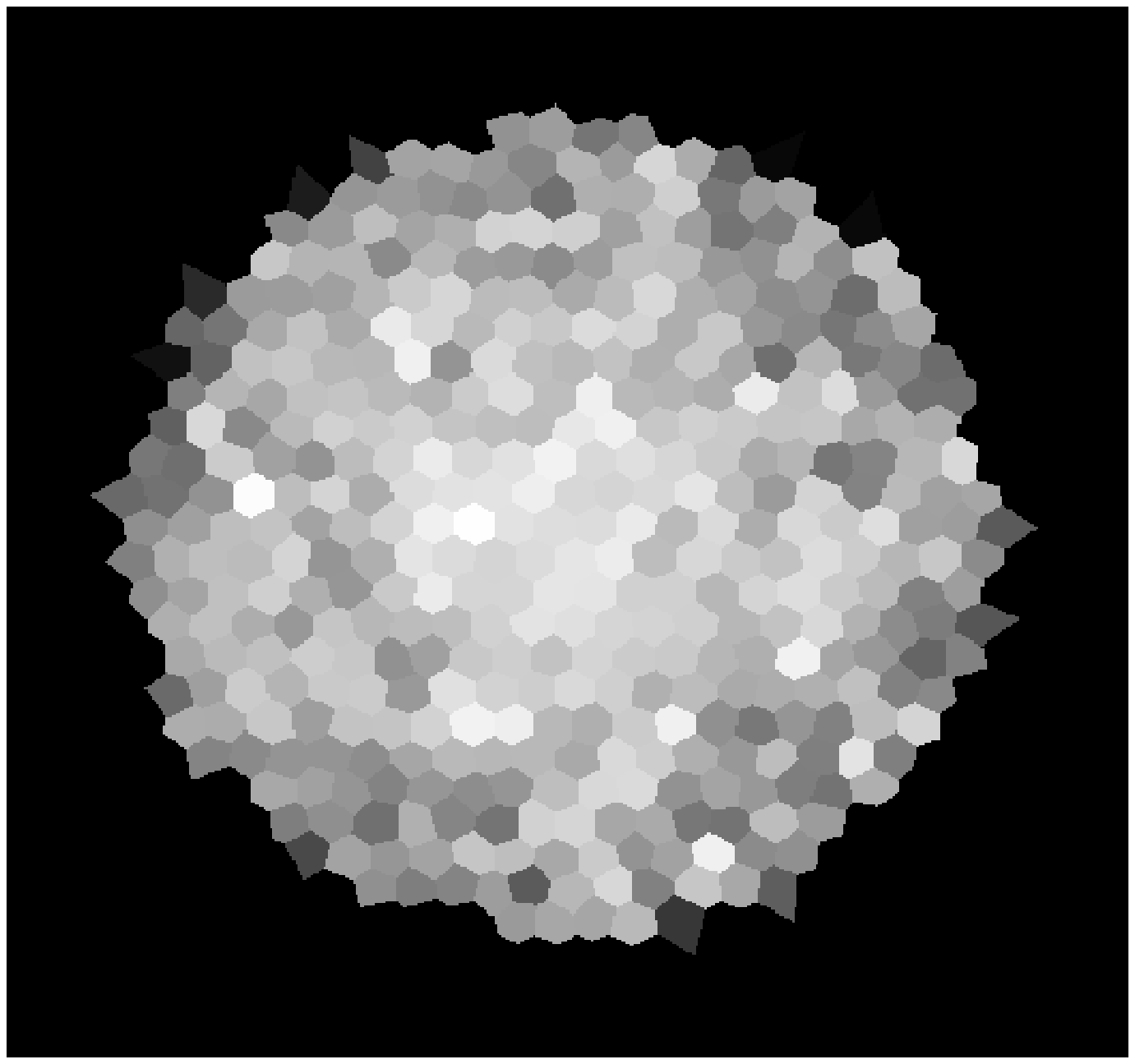} \\
   (b) \\
   \vspace{0.5cm}

   \caption{Distribution of local order for the simulated retina by
   using the shortest distance (a) and the hexagonality index(b).  The
   Voronoi cells belonging to the borders of the structure have been
   disconsidered.~\label{fig:ret_ord}}

\end{center}
\end{figure}

Another interesting situation is characterized by the progressive
variation, along space, of the spatial order.  A typical example of
such systems is the spatial distribution of photoreceptors in the
mammals retina.  Higher spatial order, actually almost perfectly
hexagonal, is observed for those receptors near the central part of
the retina, becoming less and less ordered as one moves toward the
retina periphery.  Figure~\ref{fig:ret_ord} illustrates this type of
point system, derived from a hexagonal lattice by adding perturbation
values increasing linearly with the distance from the center of the
system of points.

\begin{figure*}[h]
 \begin{center} 
   \includegraphics[scale=0.65,angle=0]{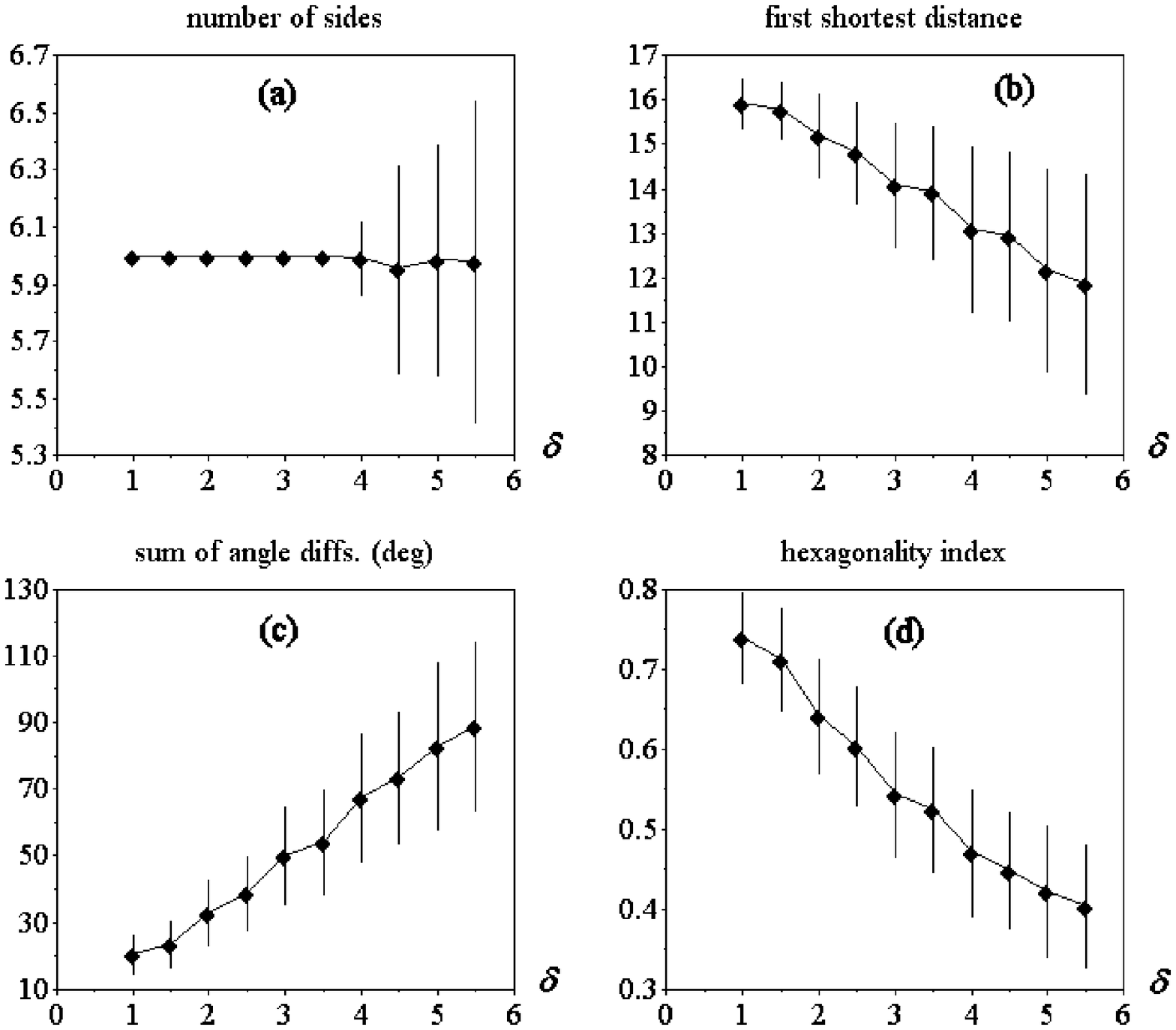}  \\
   \vspace{0.5cm}

   \caption{The average $\pm$ standard deviation of the number of
   Voronoi cell sides (a), first shortest distances (b), average of
   angle differences (c), and hexagonality indices (d), in terms of
   the perturbation intensity $\delta$ considering all individual
   points in the simulations.~\label{fig:graphs_local}}

\end{center}
\end{figure*}

Note that the conformity ratio and coefficient of variation can not be
used for quantifying local order, as the standard deviation of the
shortest distances for a single point becomes equal to zero.
Therefore, we limit the subsequent analysis to the hexagonality index,
the sum of angle differences, and the first to third shortest
distances.  Figure~\ref{fig:graphs_local} shows the average $\pm$
standard deviation considering all individual points in each of the 50
realizations of systems of points for each perturbation intensity.
Except for the higher dispersions, these results are similar to those
obtained by averaging globally over each realization shown in
Figure~\ref{fig:graphs}.

Figure~\ref{fig:local_dom} presents the results of local order
identification of the domains in Figure~\ref{fig:retseg}(a) by using
the first (a), second (b) and third (c) shortest distances as well as
the hexagonality index (d).  In order to obtain the results in
Figure~\ref{fig:local_dom}, the above measurements were obtained for
each point and Bayesian decision theory
(e.g.~\cite{CostaCesar:2001,Duda_Hart:2001}) was then applied in order
to obtain the best threshold for separating between the two groups.
The points identified as belonging to the higher order group are
represented as the squares with holes in Figure~\ref{fig:local_dom}.
Two particularly interesting results can be inferred from the obtained
results: (i) the use of successive distances tend to reduce the
correct identification of the regions and (ii) the hexagonality index
provided the best overall classification of points.  The former effect
suggests further insights about the fact that the consideration of
further shortest distances does not tend to contribute significantly
to the characterization of global order (see
Section~\ref{sec:agouti}).


Figure~\ref{fig:ret_ord} presents the quantification of the local
spatial order of the simulated retina by using the shortest distance
(a) and sum of difference angles (b).  The enhanced sensitivity of the
latter measurement is again corroborated by the more clearly defined
gradient of hexagonality values as one moves radially from the center
of the structure towards its border.  The higher uniformity of the
spatial order of the central cells is also more clearly identifiable
by using the sum of angle differences.

\section{Concluding Remarks}

The important problem of quantifying the spatial order of systems of
points has been addressed.  A series of requirements expected from a
good measurement of spatial order were identified, allowing the
discussion and comparison of a series of traditional (number of
neighbors, nearest distances between pairs of points, conformity ratio
and coefficient of variation of nearest distances) as well as a
recently introduced (the hexagonality index) and a novel (the sum of
angle differences) measurements.  

The potential of such measurements has been investigated with respect
to the quantification of global and local spatial order.  In the
former case, which involves assigning a single measurement to the
whole system of points in order to characterize its order, the
measurements were first compared with respect to simulated data,
namely hexagonal lattices with progressive perturbation intensities
$\delta$.  The obtained results allowed the objective characterization
of the linearity, sensitivity and discriminative power of each
considered measurement with respect to different levels of spatial
order and hexagonality.  The hexagonality index was verified to
account for good linearity and slightly enhanced sensitivity for
higher spatial orders, with a nearly constant discriminative power.
The sum of angle differences also yielded interesting properties
including good linearity and discriminative power.  The identification
of the type of retinal mosaics in the agouti was also considered as a
practical problem involving the estimation of spatial order.  Because
such systems involve relatively low spatial order, the sum of angle
differences was used instead of the hexagonality index.  Conformity
ratios and coefficient of variations involving the first to fifth
nearest distances were also calculated and used as subsidy for
distinguishing between the two types of mosaics.  Discriminant
analysis shown that the sum of angle differences was the only feature
allowing perfect classification.

The possibility to use the considered measurements to quantify local
spatial order was also addressed with respect to progressively
perturbed hexagonal lattices, yielding similar results as for the
global analysis, but with substantially higher dispersions of
measurement values for each perturbation intensity.  Two important
applications of local order restimation, namely the identification of
regions and analysis of systems of points involving gradients of
spatial order, were also addressed.  Such applications further
corroborated the good features of the sum of angle differences and
hexagonality index as estimators of spatial organization.

Such results imply that representative new information and insights
about ordered biological systems can be obtained by revisiting
previous investigations reported in the literature while considering a
combination of sum of angle differences and hexagonality index
measurements.  Future works should include the extension of the
angular polygonality measurements to higher dimensions and
applications to the characterization of texture and gene expression
patterns.

\vspace{1cm}

Luciano da F. Costa is grateful to FAPESP (process 99/12765-2), CNPq
(308231/03-1) and the Human Frontier Science Program for financial
support.

\bibliographystyle{plain}
\bibliography{hex_pre}

\end{document}